\documentclass[9pt,
reprint,
superscriptaddress,
showkeys,
amsmath,amssymb
]{revtex4-2}
\usepackage{xcolor}
\usepackage{xfrac}
\usepackage{graphicx}
\usepackage{dcolumn}
\usepackage{bm}
\usepackage{amsmath}
\renewcommand{\eqref}[1]{Eq.~(\ref{#1})}

\begin{document}

\title{Soliton crystal formation in Kerr cavities with an avoided mode crossing: a theoretical study}%

\author{Carlo Silvestri}
\affiliation{Institute of Photonics and Optical Science (IPOS), School of Physics, The University of Sydney, NSW 2006, Australia}
\affiliation{ARC Centre of Excellence for Optical Microcombs for Breakthrough Science (COMBS), School of Physics, The University of Sydney, NSW 2006, Australia}

\author{Caitlin E. Murray}
\affiliation{ARC Centre of Excellence for Optical Microcombs for Breakthrough Science (COMBS), School of Physics, The University of Sydney, NSW 2006, Australia}
\affiliation{Photonic Communications Lab, Department of Electrical and Computer Systems Engineering, Monash University, Clayton, VIC 3800, Australia}

\author{Chawaphon Prayoonyong}
\affiliation{ARC Centre of Excellence for Optical Microcombs for Breakthrough Science (COMBS), School of Physics, The University of Sydney, NSW 2006, Australia}
\affiliation{Photonic Communications Lab, Department of Electrical and Computer Systems Engineering, Monash University, Clayton, VIC 3800, Australia}

\author{St\'ephane Coen}
\affiliation{Department of Physics, University of Auckland, Auckland 1010, New Zealand}
\affiliation{The Dodd-Walls Centre for Photonic and Quantum Technologies, Dunedin, New Zealand}

\author{Bill Corcoran}
\affiliation{ARC Centre of Excellence for Optical Microcombs for Breakthrough Science (COMBS), School of Physics, The University of Sydney, NSW 2006, Australia}
\affiliation{Photonic Communications Lab, Department of Electrical and Computer Systems Engineering, Monash University, Clayton, VIC 3800, Australia}

\author{C. Martijn de Sterke}
\affiliation{Institute of Photonics and Optical Science (IPOS), School of Physics, The University of Sydney, NSW 2006, Australia}
\affiliation{ARC Centre of Excellence for Optical Microcombs for Breakthrough Science (COMBS), School of Physics, The University of Sydney, NSW 2006, Australia}

\author{Antoine F. J. Runge}
\affiliation{Institute of Photonics and Optical Science (IPOS), School of Physics, The University of Sydney, NSW 2006, Australia}
\affiliation{ARC Centre of Excellence for Optical Microcombs for Breakthrough Science (COMBS), School of Physics, The University of Sydney, NSW 2006, Australia}

\begin{abstract}
We theoretically and numerically investigate the formation of soliton crystals in Kerr microresonators in the presence of an avoided mode crossing (AMX). Our study combines dynamical simulations based on a modified Lugiato–Lefever equation (LLE) with a stability analysis of its stationary solutions. We show that, depending on its strength and spectral position, the AMX can either stabilize otherwise unstable soliton crystals or induce Turing patterns which subsequently seeds soliton crystal formation. Both perfect and imperfect soliton crystals can form below the pump threshold for spatiotemporal chaos, and we identify the conditions required for perfect crystals. Finally, we investigate modulation instability in the presence of an AMX, showing that it modifies the parametric gain and can suppress or promote Turing pattern formation depending on its spectral position.
\end{abstract}

\maketitle


\section{Introduction}
Optical frequency combs have spectra with equally spaced optical lines and have reshaped fundamental science by enabling frequency measurements with unprecedented precision~\cite{Udem_2002,Hansch}. Originally developed around bulky laser sources, frequency combs can now be generated in compact passive microresonators externally driven by a continuous-wave (CW) optical pump field~\cite{DelHaye_2007,Kippenberg_2018}. These \textit{microcombs} are relevant for applications such as spectroscopy, telecommunications, and sensing~\cite{Pasquazi_2018, Sun2023Mar}. The generation of microcombs is often associated with the formation of cavity solitons (CSs), localized pulses sustained by the double balance between gain and loss, as well as Kerr nonlinearity and anomalous dispersion~\cite{Kippenberg_2018,Pasquazi_2018}. The formation of CSs in passive cavities is successfully modeled using the Lugiato–Lefever equation (LLE)~\cite{LLE_1987,Coen_2013,Coen2_2013,Pasquazi_2018}.

It was recently shown that CSs can combine to form a {\sl soliton crystal} state~\cite{Cole_2017,Karpov_2019}. Bound soliton states, of which soliton crystals are a particular class, arise when the CSs have oscillatory tails, that cause the CSs to take discrete temporal separations~\cite{Wang:17}. In microresonators this is usually provided by the presence of avoided mode crossings (AMXs)~\cite{Cole_2017,HerrPRL2014, Wang2018May}. 

Briefly, an AMX is a spectrally localized perturbation of the microresonator dispersion, which modifies the CS spectrum and leads to the emergence of an effective secondary pump. This interferes with the primary driving field, generating a modulated intracavity background that enables the CSs to arrange into a stable soliton crystal~\cite{Cole_2017,Karpov_2019,Wang:17}. Other mechanisms that can induce bound soliton states in Kerr cavities include high-order dispersion, Kelly sidebands, and multi-frequency pumping~\cite{Silvestri_2025,Wang:17,taheri2017optical,Lu_2021}. 

Soliton crystals can include one or more defects, such as vacancies or dislocations, and in this case have comb spectra with a spacing equal to the free spectral range (FSR)~\cite{Cole_2017}. 
However, the solitons can also be uniformly spaced, forming a perfect soliton crystal (PSC)~\cite{Karpov_2019}. The spectrum of a PSC corresponds to a frequency comb with lines separated by an integer multiple $n$ of the cavity FSR, where $n$ is the number of solitons in the crystal. Karpov \textit{et al.} showed that in a microresonator with quadratic dispersion and AMX, detuning scans at fixed pump power lead to deterministic PSC formation only below a threshold corresponding to the lower boundary of the spatiotemporal chaos region in the LLE phase diagram~\cite{Karpov_2019,Leo_2013}. 

Despite the importance of AMX for the formation of soliton crystals in microresonators, its precise role and the mechanism underlying soliton crystal formation remain unclear. Most reports are primarily experimental, with theoretical work largely limited to dynamical simulations validating specific experimental results~\cite{Karpov_2019,Cole_2017}. Therefore, a comprehensive theoretical understanding is still lacking.

In this work, we develop this understanding by adopting a modified LLE that incorporates the effect of the AMX and using it to investigate the formation of soliton crystals. Our analysis combines dynamical simulations with a stability analysis of the LLE stationary soliton-crystal and Turing-pattern solutions. Our results show that an AMX can produce distinct effects: (i) they can stabilize soliton-crystal states that would be unstable in the absence of the AMX; (ii) they can stabilize Turing patterns or modify their periodicity through a pinning mechanism, after which these patterns act as seeds for soliton-crystal formation; and (iii) they can alter the parametric gain spectrum, thereby inducing the formation of Turing patterns different from those supported without the AMX. Finally, we show that both perfect and imperfect soliton crystals can form below the threshold for spatiotemporal chaos, depending on the spectral position of the AMX. 

This paper is structured as follows: in Section~\ref{sec_model}, we introduce the theoretical model used in this work; in Section~\ref{secresults}, we present the numerical and theoretical results; in Section~\ref{sec_concl}, we draw the conclusions of the work; in Appendix~\ref{MI_sec}, we perform the MI analysis in the presence of AMX, deriving the parametric gain associated with CW destabilization.

\section{Theoretical model}\label{sec_model}
We begin by considering the dynamics of the electric field envelope $E(t,\tau)$ in a high-$Q$ ring microresonator, described by the normalized LLE~\cite{Coen2_2013,Pasquazi_2018}:
\begin{equation}
\frac{\partial E(t,\tau)}{\partial t} = \left[-1 + i\left(|E|^2 - \Delta\right) - i\eta\frac{\partial^2}{\partial \tau^2}\right]E + S,
\label{eq:GLLE}
\end{equation}
where $\eta = \mathrm{sign}(\beta_2)$, with $\beta_2$ the $2$nd-order dispersion coefficient, $t$ denotes the slow time, $\tau$ the fast time, $\Delta$ the normalized detuning, and $S$ the normalized driving field amplitude. We take the dispersion to be anomalous, and therefore $\eta = -1$.

We implement the AMX in the LLE as a $\delta$-function perturbation of the dispersion relation~\cite{Cole_2017,Karpov_2019}. The resulting dispersion relation reads:
\begin{equation}
\beta(\omega)
=
\eta \omega^2
+
\Delta_{\mathrm{AMX}}
\,\delta(\omega-\omega_{\mathrm{AMX}}).
\label{eq:dis_rel}
\end{equation}
where $\Delta_{\mathrm{AMX}}$ denotes the AMX strength, and $\omega_{\mathrm{AMX}}$ is the angular frequency at which the AMX occurs. The equivalent expression in the time domain can be obtained by computing the inverse Fourier transform of $\beta(\omega)\hat{E}(\omega)$:
\begin{align}
\mathcal{F}^{-1}\{\beta(\omega)\hat{E}(\omega)\}
=
-\eta \frac{\partial^2 E}{\partial \tau^2}
+
\Delta_{\mathrm{AMX}}
\hat{E}(\omega_{\mathrm{AMX}})
e^{-i\omega_{\mathrm{AMX}}\tau},
\label{eq:FT}
\end{align}
The first term on the right-hand side of \eqref{eq:FT} arises from the parabolic dispersion term in \eqref{eq:dis_rel}, while the second one is associated with the AMX. Therefore, the AMX can be taken into account by modifying the LLE \eqref{eq:GLLE} as:
\begin{equation}
\begin{split}
\frac{\partial E}{\partial t}
&=
\underbrace{
\left[
-1 - i\Delta
- i\eta \frac{\partial^2}{\partial \tau^2}
+ i|E|^2
\right]E
+S
}_{\text{Conventional LLE terms}}
\\
&\hspace{2cm}
+
\underbrace{
i\Delta_{\mathrm{AMX}}
\hat{E}(\omega_{\mathrm{AMX}})
e^{-i\omega_{\mathrm{AMX}}\tau}
}_{\text{AMX}} .
\end{split}
\label{LLE_AMX}
\end{equation}
The AMX term thus acts as an additional effective pump, naturally emerging from the introduction of a $\delta$-like perturbation in the dispersion relation. Its spectral position is fixed at $\omega_{\mathrm{AMX}}$, while its amplitude is proportional to both the AMX strength and the amplitude of the spectral component of the intracavity field at $\omega_{\mathrm{AMX}}$.
\begin{figure}[t!]
\centering
\includegraphics[width=0.5\textwidth]{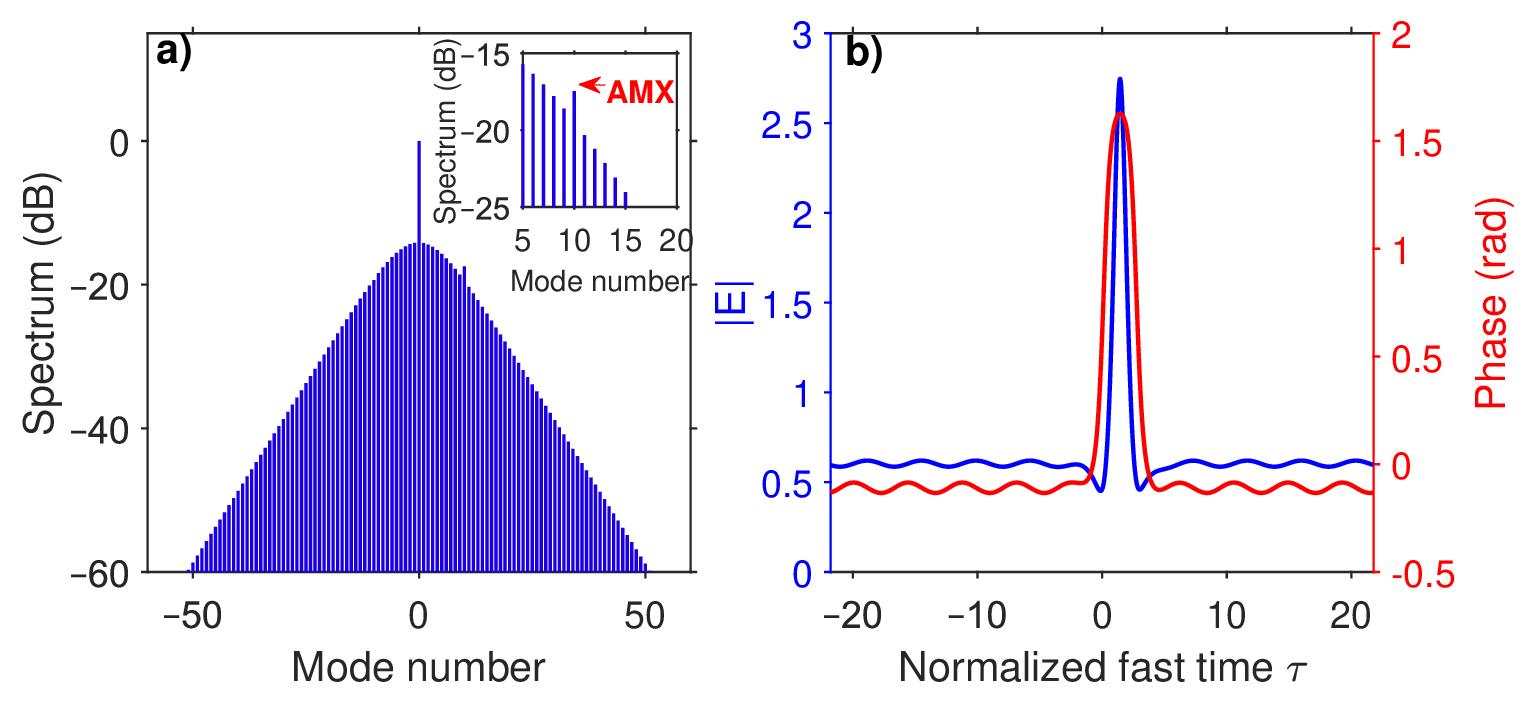}
\vskip -1mm
\caption{{Single soliton in the presence of an AMX at mode 10, obtained by solving \eqref{LLE_AMX} with $\Delta_\mathrm{AMX}=1$, $\Delta=3.3$, and $X=3.5$. (a) Optical spectrum, showing enhanced power at the AMX frequency; the inset provides a zoom around the AMX mode. (b) Temporal profiles of the intracavity field amplitude (blue) and phase (red).}}
\label{fig0}
\end{figure}
\begin{figure*}[t]
\centering
\includegraphics[width=1\textwidth]{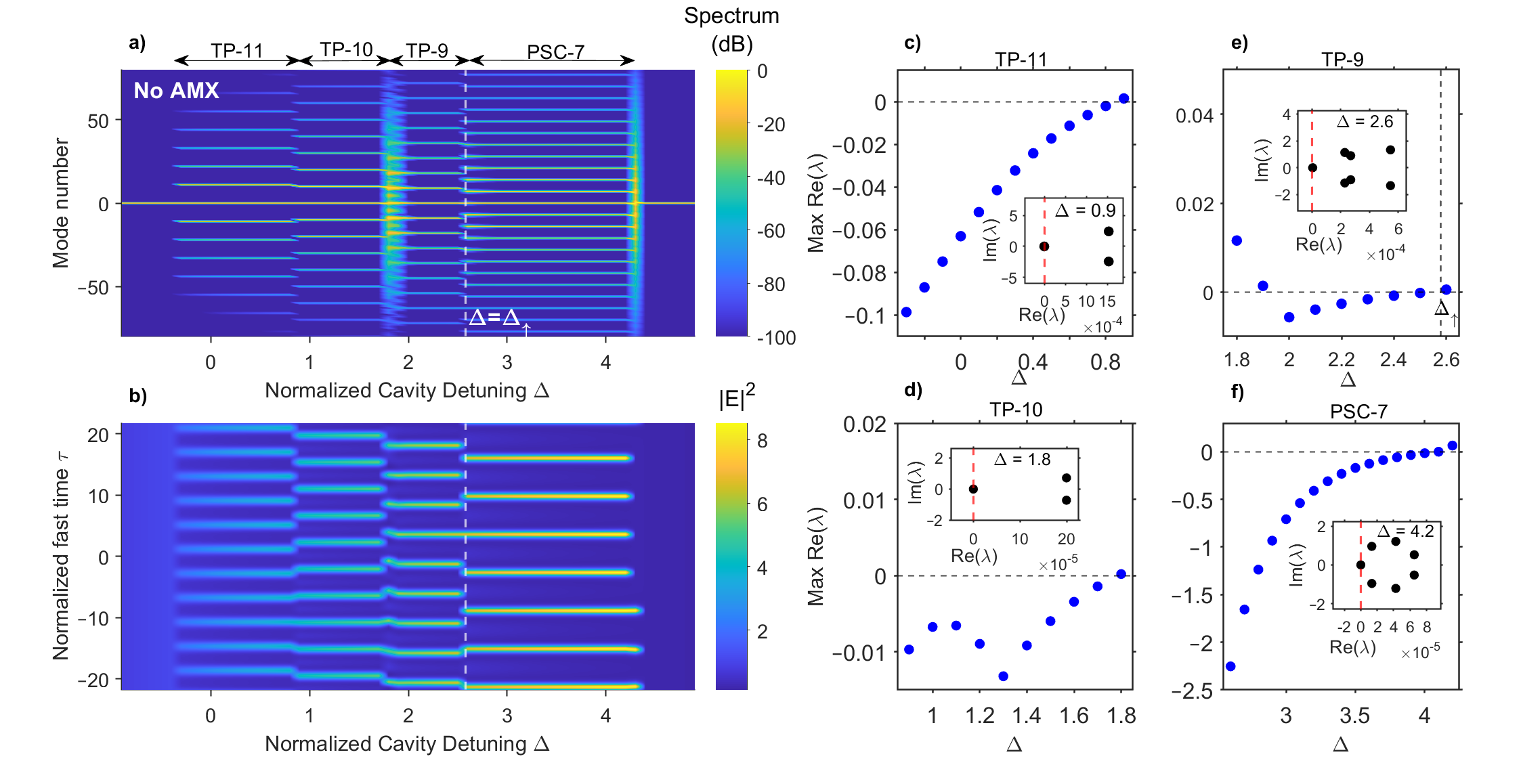}
\vskip -1mm
\caption{Detuning scan in the absence of an AMX. Evolution of (a) the optical spectrum and (b) the intracavity intensity profile as a function of the normalized detuning $\Delta$. Dashed white vertical lines indicate the upswitching detuning $\Delta_\uparrow=2.58$. (c)-(f) Maximum real part of the temporal eigenvalues versus $\Delta$ for the Turing patterns (c) TP-11, (d) TP-10, (e) TP-9, and, (f) for the perfect soliton crystal PSC-7. The insets show the eigenvalue configuration at the onset of instability.} 
\label{fig1}
\end{figure*}
The optical spectrum and the temporal intracavity intensity profile of a single CS in the presence of an AMX at mode 10 are shown in Fig.~\ref{fig0}(a) and (b), respectively. These were obtained by integrating \eqref{LLE_AMX}. The optical spectrum exhibits an excess of power at the AMX frequency, which effectively acts as an additional coherent driving field (Fig.~\ref{fig0}(a)).  
In time, both the amplitude and phase of the intracavity field develop a periodic modulation with a frequency matching that of the AMX (Fig.~\ref{fig0}(b)). Phase and amplitude modulations are known to trap cavity solitons at specific positions~\cite{ErkintaloPhase,Wang:17}: e.g. solitons drift toward phase maxima and away from phase minima, making the maxima stable equilibrium positions~\cite{Maggipinto,ErkintaloPhase}. The AMX therefore creates a set of preferred trapping sites, enabling the formation of soliton crystals~\cite{Cole_2017,Wang:17}.

Note that the AMX term in \eqref{LLE_AMX} depends on the entire temporal intracavity field profile $E\left(\tau\right)$, rendering the AMX term nonlocal. Also, when the intracavity field is in a CW state and $\omega_{\mathrm{AMX}} \neq 0$, then $\hat{E}(\omega_{\mathrm{AMX}})=0$, and the AMX does not contribute.

\section{Results}\label{secresults}
\subsection{Numerical and theoretical procedure}\label{sec_procedure}
In the presence of transient or spatiotemporal chaos, PSCs do not form deterministically~\cite{Karpov_2019}. Therefore, we set the pump power $X= |S|^2 =3.5$, which lies below the thresholds for both spatiotemporal chaos ($X_{\mathrm{th,STC}}\approx9$) and transient chaos ($X_{\mathrm{th,TC}}\approx16$)~\cite{Leo_2013,Karpov_2019}.

Our study is based on two complementary theoretical approaches. The first consists of dynamical simulations of \eqref{LLE_AMX}, performed using a sixth-order split-step Fourier method~\cite{BLANES2002313}. The second consists of calculating the stationary solutions of \eqref{LLE_AMX} and performing their linear stability analysis. In this section we will present the results obtained with both methods and discuss their consistency throughout the analysis. We now briefly describe these two approaches.

In experiments, CSs are typically generated by scanning the driving-field frequency across a cavity resonance at fixed pump power~\cite{Kippenberg_2018}. We adopt the same approach in our dynamical simulations, i.e. we sweep the normalized detuning $\Delta$ at fixed $X$. Each dynamical simulation at fixed detuning spans at least 20 million roundtrips. Near the onset of instabilities of the Turing patterns and the PSC solutions, simulations are extended up to a billion roundtrips to capture destabilization occurring on very slow timescales. 

We also compute the stationary solutions of~\eqref{LLE_AMX} using a Newton method and analyze their stability by calculating the temporal eigenvalues $\lambda$ of the corresponding Jacobian matrix~\cite{NumRec}. Stability depends on the sign of the real parts of these eigenvalues~\cite{Coen2_2013}. If one of the eigenvalues has a positive real part the associated solution is unstable; the stability is thus determined by the eigenvalue with the largest real part~\cite{NumRec}.

We distinguish Turing rolls from PSCs by injecting a phase pulse into the cavity which aims to erase one of the pulses of the pattern solution~\cite{Jang_2015,Bartolo:22, Barland_2002}. In the case of a TP, the erased pulse reforms, revealing the global nature of the pattern. By contrast, for a PSC the pulse is permanently erased, demonstrating the localized nature of each pulse composing the crystal~\cite{Bartolo:22}. 

For reference, we first consider the case without AMX, i.e., $\Delta_{\mathrm{AMX}}=0$. Then, in Sec.~\ref{subsec_noAMX}, we introduce the AMX to assess how it modifies the dynamics and stability scenario. We use parameters representative of a $\rm{MgF_2}$ resonator with a free spectral range of $35.2~{\rm GHz}$, quality factor $Q\approx4\times10^8$, and finesse $\mathcal{F}=73\times10^3$~\cite{Herr_2014,Pasquazi_2018}. For our parameters, the normalized roundtrip time is $\tau_\mathrm{RT}=43.6$~\cite{Pasquazi_2018}.

\subsection{The reference case: no AMX}\label{subsec_noAMX}
Figures~\ref{fig1}(a)-(b) show the evolution of intracavity spectral and temporal intensity profiles, respectively, as the normalized detuning $\Delta$ is swept between $-0.9$ and $4.9$ with step $0.1$. These results were obtained by solving \eqref{LLE_AMX} with $\Delta_{\mathrm{AMX}}=0$. At $\Delta = -0.3$, the CW state becomes unstable, leading to the formation of a Turing pattern with 11 rolls (TP-11) via modulational instability (MI). This is consistent with a MI analysis, as for these parameters mode 11 is the closest to the parametric gain peak (see Appendix~\ref{MI_sec}). The TP-11 state persists up to $\Delta = 0.9$, where it destabilizes and changes to a Turing pattern with 10 rolls (TP-10).

To investigate the stability of the TP-11 state, we show in Fig.~\ref{fig1}(c) the largest real part among the temporal eigenvalues, i.e. $\max \mathrm{Re}(\lambda)$, as a function of $\Delta$ over the detuning range in which the TP-11 solution is observed in the dynamical simulations. Note that the solution becomes unstable at $\Delta\approx 0.9$, where $\max \mathrm{Re}(\lambda)$ becomes positive. The inset of Fig.~\ref{fig1}(c) shows the corresponding eigenvalue spectrum, indicating that the destabilization occurs through a Hopf bifurcation, with a pair of complex-conjugate eigenvalues crossing into the $\mathrm{Re}(\lambda)>0$ half-plane. This suggests that the destabilization is associated with an Eckhaus instability~\cite{Perinet2017,Gomila2022}. 

The TP-10 state is stable between $\Delta=0.9$ and $\Delta=1.8$, where it also undergoes a Hopf bifurcation (Fig.~\ref{fig1}(d)) and evolves into a Turing pattern with 9 rolls (TP-9, see Fig.~\ref{fig1}(a)-(b)). We attribute this transition also to an Eckhaus instability~\cite{Perinet2017}.
The TP-9 state initially exhibits slow-time oscillations up to $\Delta \approx 1.9$, beyond which the oscillations disappear (Fig.~\ref{fig1}(a)–(b)). The regime in which the Turing rolls oscillate corresponds to the dense-spectrum region between $\Delta=1.8$ and $1.9$ in Fig.~\ref{fig1}(a). 
\begin{figure}[t!]
\centering
\includegraphics[width=0.5\textwidth]{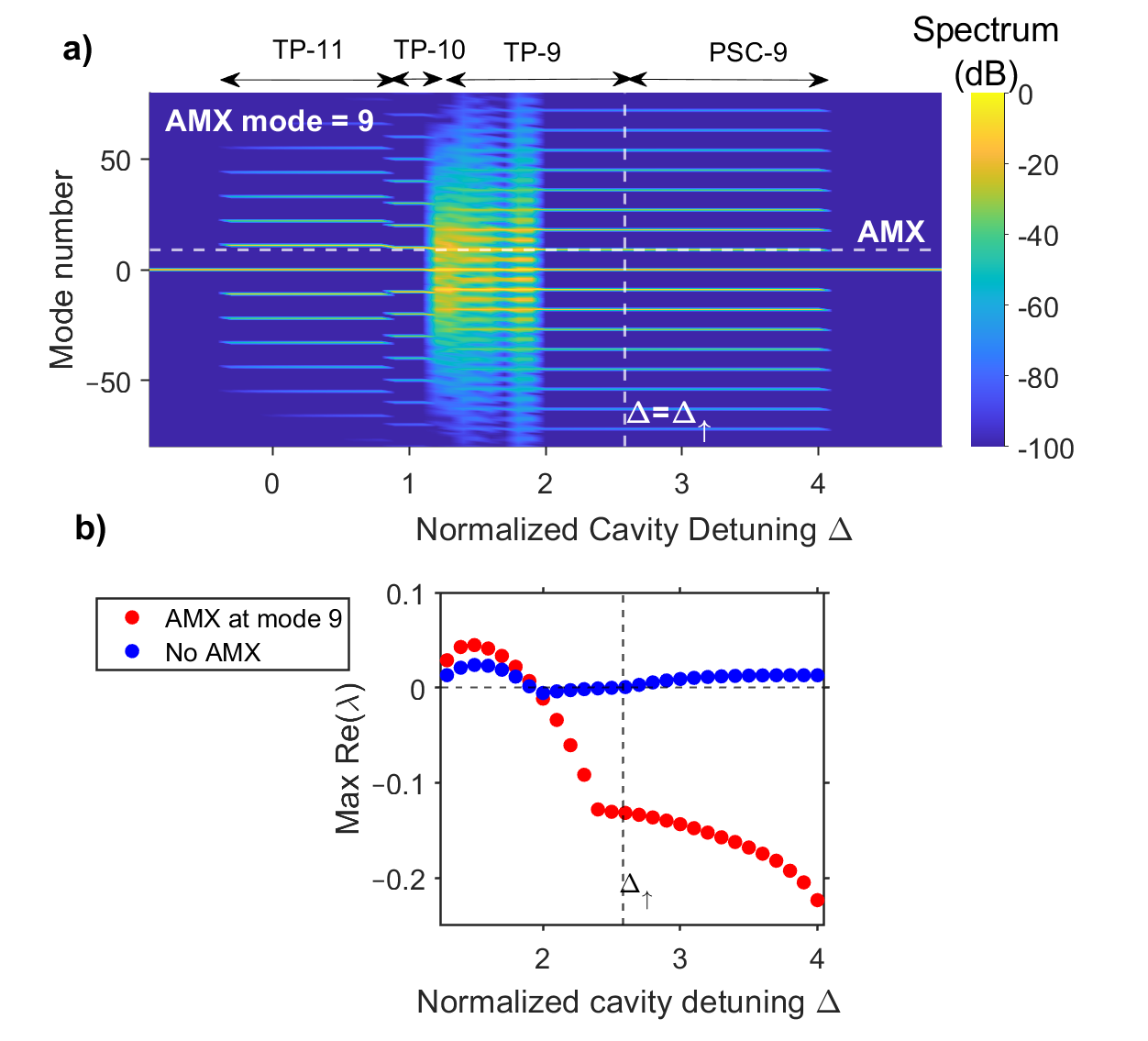}
\vskip -1mm
\caption{Stabilization of an otherwise unstable PSC, with an AMX at mode 9. (a) Evolution of the optical spectrum as a function of detuning in the presence of an AMX positioned at mode 9. (b) Maximum real part of the temporal eigenvalues, $\mathrm{max}$ $\mathrm{Re}(\lambda)$, as a function of detuning for the 9-pulse states (TP-9 and PSC-9), with (red) and without (blue) AMX. }
\label{fig2}
\end{figure}

Just above the up-switching point, $\Delta_{\uparrow}=2.58$, the system enters the CW bistability region, where CSs can exist. At that point, we observe that the TP-9 state destabilizes, evolving into a perfect soliton crystal with different periodicity (PSC-7). The dynamical simulations are in excellent agreement with the stability analysis of the TP-9 state, as shown in Fig.~\ref{fig1}(e). The solution is initially unstable at $\Delta=1.8$ and $1.9$. In the dynamical simulations, this instability manifests as temporal oscillations and a dense optical spectrum (Fig.~\ref{fig1}(a)-(b)). It then is stable up to $\Delta=2.5$, before destabilizing again at $\Delta=2.6$, where multiple eigenvalues acquire a positive real part (inset of Fig.~\ref{fig1}(e)) and the dynamical simulations show the formation of the PSC-7 state. The scenario described in Fig.~\ref{fig1} is consistent with results presented in \cite{Qi2019Sep}. Note that all Turing patterns arising during the detuning scan, namely TP-11, TP-10, and TP-9, destabilize and disappear before or near $\Delta_\uparrow$, preventing the formation of PSCs with the corresponding periodicities (namely PSC-11, PSC-10, and PSC-9). As we will show in Sec.~\ref{sec_stab}, the presence of an AMX stabilizes these patterns over a broader detuning range, allowing the system to evolve toward PSCs states with matching periodicity.
\begin{figure}[t!]
\centering
\includegraphics[width=0.5\textwidth]{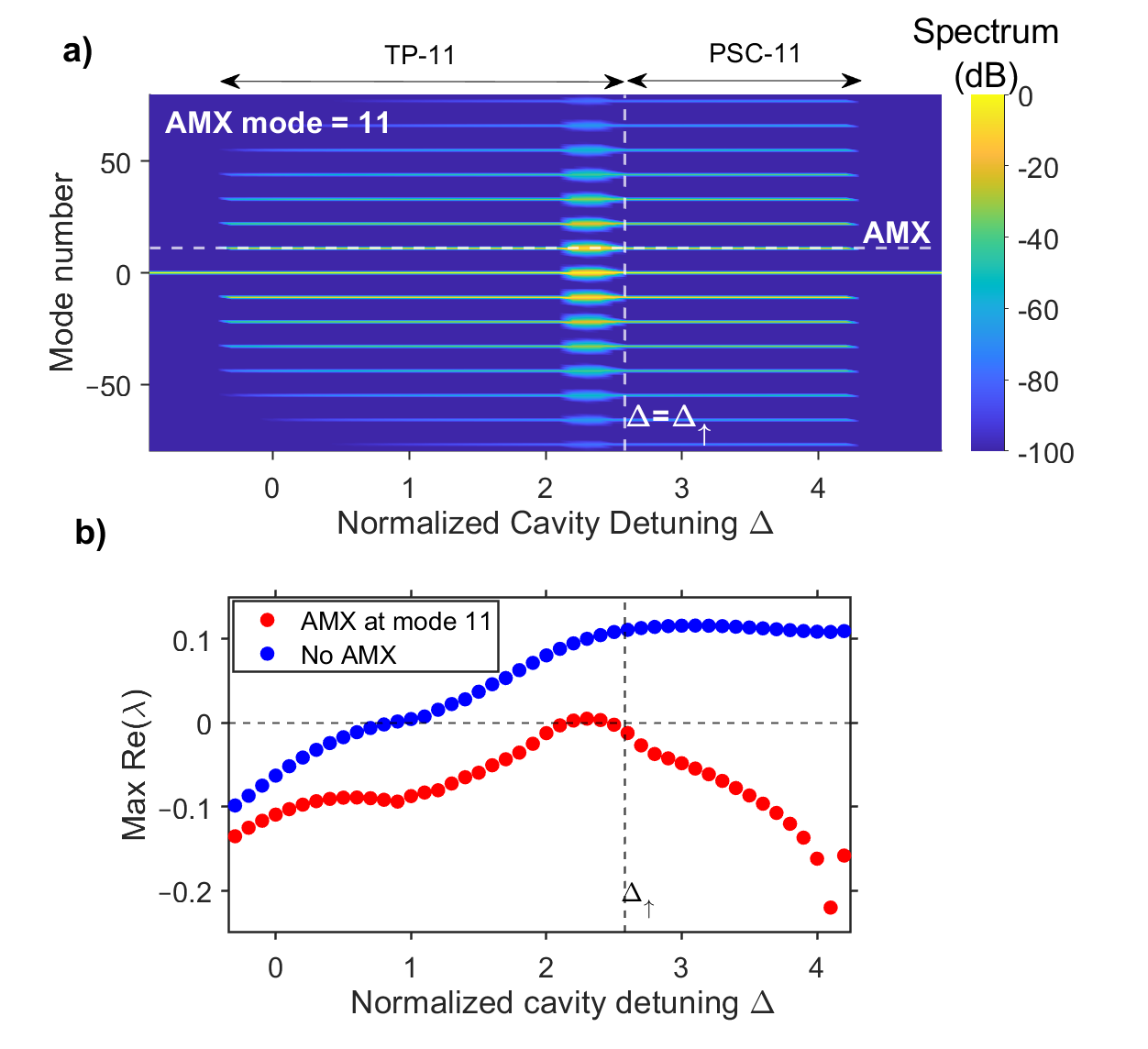}
\vskip -1mm
\caption{Stabilization of an otherwise unstable Turing pattern with an AMX at mode 11, leading in turn to the formation of a PSC. (a) Evolution of the optical spectrum versus $\Delta$ in the presence of an AMX positioned at mode 11. (b) $\mathrm{Max}$ $\mathrm{Re}(\lambda)$ versus $\Delta$ for the 11-pulse states (TP-11 and PSC-11), with (red) and without (blue) AMX.}
\label{fig3}
\end{figure}
Nevertheless, we find that a stable PSC with 7 solitons can still emerge without AMX in the standard LLE. A detailed investigation of PSC formation in the absence of AMX is deferred to future work.

This PSC-7 remains stable in the dynamical simulations up to $\Delta=4.2$ (Fig.~\ref{fig1}(a)-(b)), where it destabilizes into three non-equally spaced solitons, followed by the formation of a CW state as the detuning is further increased. The stability analysis predicts the destabilization of the PSC-7 at $\Delta=4.2$ (Fig.~\ref{fig1}(f)), in excellent agreement with the simulations.

\subsection{The stabilizing action of the AMX} \label{sec_stab}
We showed in Section~\ref{subsec_noAMX} that, in the absence of AMX, the system evolves toward the preferred Turing-pattern states TP-11, TP-10, and TP-9, depending on the detuning. However, these destabilize before or near $\Delta_\uparrow$ and therefore do not act as a seed for formation of PSCs with the same periodicity (respectively PSC-11, PSC-10, PSC-9). To see the effect of the inclusion of an AMX, we now consider the cases where an AMX is introduced at spectral positions matching the periodicities of these patterns (mode 11, 10 and 9, respectively), with strength $\Delta_\mathrm{AMX}=0.25$ in all cases.

\begin{figure}[t!]
\centering
\includegraphics[width=0.5\textwidth]{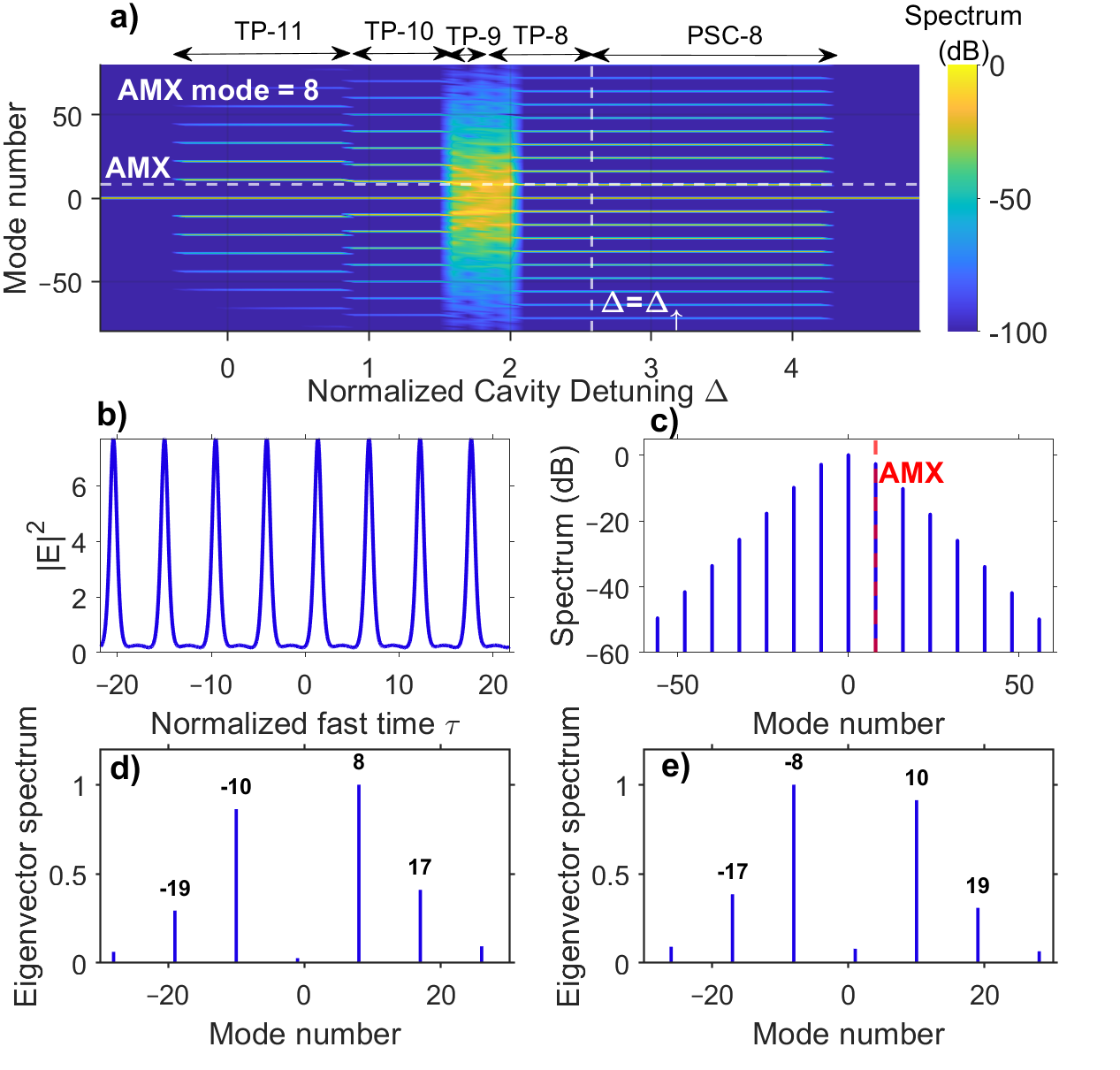}
\vskip -1mm
\caption{Pinning action leading to PSC formation. (a) Evolution of the optical spectrum as a function of detuning with the AMX positioned at mode 8. (b) Temporal and (c) spectral profiles of the PSC-8 state at $\Delta=3.3$. (d)–(e) Normalized Fourier transform of the eigenvectors (eigenvector spectra) corresponding to the two eigenvalues with the largest real part at $\Delta=1.9$, when the TP-8 state forms, exhibiting prominent peaks at modes $\pm 8$ in agreement with the observed pattern periodicity.}
\label{fig4}
\end{figure}

First, we consider the AMX positioned at mode $9$. The evolution of the optical spectrum as a function of detuning is shown in Fig.~\ref{fig2}(a). For $\Delta < \Delta_\uparrow$, the dynamics closely resemble the case without AMX shown in Fig.~\ref{fig1}, with the formation of a TP-11 state that destabilizes into TP-10, which in turn destabilizes at higher detuning into a TP-9 state. An important difference emerges, however, once the system enters the bistable region. Here, the PSC composed of nine solitons remains stable as the detuning increases, in contrast to the AMX-free case. To investigate this further, Figure~\ref{fig2}(b) shows the maximum real part of the temporal eigenvalues as a function of detuning in the presence of the AMX at mode 9 (red line), together with the corresponding results in the absence of AMX (blue); the presence of the AMX evidently stabilizes the nine-pulse stationary solution for $\Delta>\Delta_\uparrow$, leading to the formation of a stable PSC-9. Therefore, a soliton crystal that is unstable without AMX can become stable once the AMX is introduced and spectrally aligned with the fundamental mode of the crystal.
\begin{figure}[t!]
\centering
\includegraphics[width=0.5\textwidth]{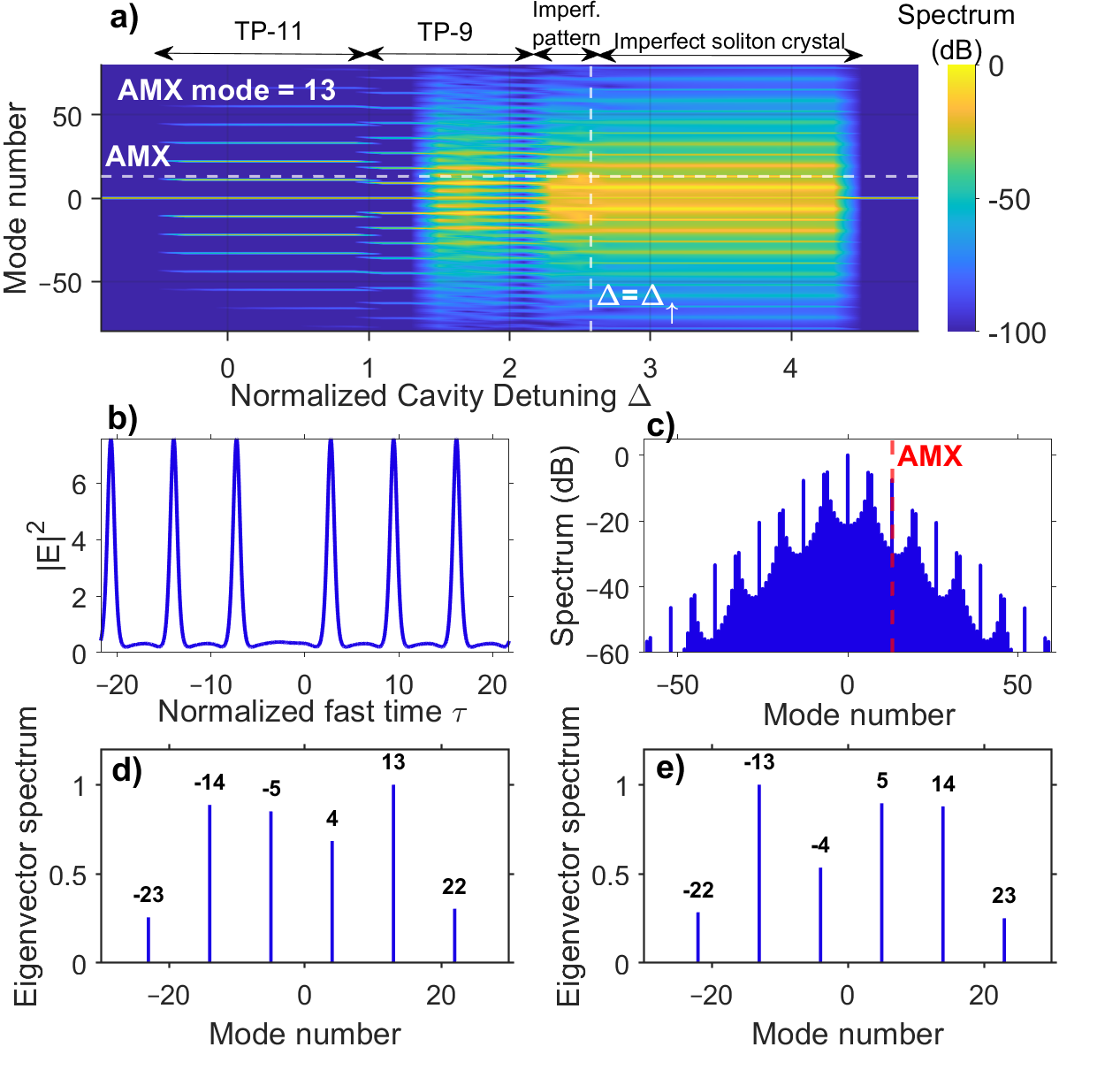}
\vskip -1mm
\caption{Pinning action leading to an imperfect soliton crystal. (a) Optical spectrum versus detuning with the AMX positioned at mode 13. (b) Temporal and (c) spectral profiles of an imperfect soliton crystal at $\Delta=3.3$. (d)–(e) Normalized Fourier transform of the eigenvectors (eigenvector spectra) corresponding to the two eigenvalues with the largest real part at $\Delta=2.3$, when the TP-9 is unstable and disappears, leading to a non-periodic Turing pattern.}
\label{fig5}
\end{figure}
Next we consider the case in which the AMX is placed at mode 11. In contrast to the AMX-free scenario, where TP-11 destabilizes at low detuning, here the TP-11 state persists across the entire range $\Delta < \Delta_{\uparrow}$ (Fig.~\ref{fig3}(a)). For $2.2 \leq \Delta \leq 2.4$, temporal oscillations of the Turing rolls lead to a denser spectrum (see Fig.~\ref{fig3}(a)). Nevertheless, upon entering the bistable region, this state seeds the formation of an eleven-soliton crystal (PSC-11), which is observed in the dynamical simulations up to $\Delta = 4.2$.
The eigenvalue analysis (Fig.~\ref{fig3}(b)) shows that the stationary solution with 11 pulses is stable up to $\Delta = 2.1$ and becomes weakly unstable, with the maximum real part of the eigenvalues remaining close to zero, in the interval $2.2 \leq \Delta \leq 2.4$, precisely where the denser spectrum is observed in Fig.~\ref{fig3}(a). The solution then remains stable for $\Delta>2.4$, and therefore corresponds to a stable PSC-11 for $\Delta>\Delta_\uparrow$. Thus, the AMX again has a stabilizing effect. Unlike the case of mode 9, however, the stabilization here occurs at the level of the Turing pattern, which subsequently enables the formation of a stable PSC.

Similar behavior occurs when the AMX is positioned at mode 10 (not shown). The system first transitions from TP-11 to TP-10, as in the case without AMX, but TP-10 no longer destabilizes and instead seeds the formation of a PSC-10 above the upswitching point.

\subsection{The pinning action of the AMX} \label{sec_pinning}
We now consider the AMX positioned at mode 8, which does not match the periodicity of any preferred Turing pattern. The dynamical simulations (Fig.~\ref{fig4}(a)) show that, following the sequence TP-11$\rightarrow$TP-10$\rightarrow$TP-9, the latter disappears at $\Delta = 1.9$ and is replaced by a Turing pattern with a periodicity of 8 FSRs (TP-8). The TP-9 state is associated with a dense spectrum in Fig.~\ref{fig4}(a) due to temporal oscillations of the nine rolls (time domain not shown here).

The resulting TP-8 state then seeds the formation of a PSC with 8 solitons for $\Delta>\Delta_\uparrow$, whose temporal profile and spectrum are shown in Fig.~\ref{fig4}(b) and (c), respectively. Here, the action of the AMX is to force the system onto a Turing pattern solution whose first adjacent mode coincides with the position of the AMX. We define the first adjacent mode as the first positive-frequency mode adjacent to the central mode (mode 0) with nonzero intensity in the optical spectrum. For a PSC-$n$ state, the first adjacent mode is separated from mode 0 by $n$~FSRs, as shown in Fig.~\ref{fig4}(c) for a PSC-8. In contrast, imperfect soliton crystals exhibit dense spectra, such that the first adjacent mode is separated from mode 0 by 1~FSR.

The PSC-8 does not appear without AMX, as shown in Fig.~\ref{fig1}, and is therefore a direct result of the AMX. To corroborate this, we examine the instability of the TP-9 state at $\Delta = 1.9$, where TP-8 emerges. We compute the Fourier transform of the eigenvectors (hereafter referred to as eigenvector spectra) associated with the two eigenvalues having the largest real part, namely the complex-conjugate pair $\lambda = 0.03 \pm 0.06 i$. These are shown in Fig.~\ref{fig4}(d)–(e) and reveal which spectral components exhibit the fastest growth as TP-9 destabilizes at $\Delta=1.9$. The most prominent peaks in these spectra occur precisely at modes $-8$ and $+8$. This indicates that the presence of the AMX selectively enhances the growth of these modes. As the spectral line at mode $+8$ increases in amplitude, the effective pump term in \eqref{LLE_AMX} becomes progressively stronger, ultimately driving the formation of the TP-8 state. We verified that the PSC-8 that forms for $\Delta>\Delta_\uparrow$ is stable.  Similar behavior is observed when the AMX is positioned at mode 7; it leads to a TP-7 state, which then seeds the generation of a PSC with the same periodicity.

For completeness, we also consider the AMX positioned at mode $13$. The simulations in Fig.~\ref{fig5}(a) show that, when TP-9 disappears at $\Delta = 2.3$, mode 13 increases in intensity in the optical spectrum but does not become the first adjacent mode. As a result, in the range between $\Delta = 2.3$ and $\Delta_\uparrow$, a dense-spectrum pattern associated with non-equally spaced rolls (imperfect pattern) is formed. For $\Delta > \Delta_\uparrow$, this state evolves into an imperfect soliton crystal, with a temporal profile consisting of six irregularly spaced solitons (see Fig.~\ref{fig5}(b)). Its optical spectrum exhibits a peak at the AMX frequency, but remains dense, with the most prominent peak occurring at mode 6 (see Fig.~\ref{fig5}(c)). The eigenvectors corresponding to the eigenvalues with the largest real part at $\Delta = 2.3$ now exhibit prominent peaks at mode 13, a clear fingerprint of the AMX action (see Fig.~\ref{fig5}(d)–(e)). Unlike the previous cases, here the AMX does not generate a perfect soliton crystal with periodicity 13 though it induces a spectral peak at this mode. We attribute this to the reduced intensity of the effective pump associated with the AMX, since it depends on $\hat{E}\left(\omega_\mathrm{AMX}\right)$, i.e. the spectral intensity at the AMX frequency (see \eqref{LLE_AMX}). This intensity decreases away from the center of the spectrum, resulting in a progressively weaker effective pump. Consequently, for a fixed AMX strength $\Delta_\mathrm{AMX}$, the case with the AMX at mode 13 corresponds to a weaker effective pump than the case at mode 8, preventing mode 13 from becoming dominant. This result also shows that, for $X$ below the pump threshold for spatiotemporal chaos $X_\mathrm{th,STC}$, imperfect soliton crystals can form, indicating that this region of parameter space is not exclusively associated with the deterministic formation of PSCs.
\subsection{Perfect and imperfect soliton crystals} \label{sec_perf_imperf}
\begin{figure}[t!]
\centering
\includegraphics[width=0.5\textwidth]{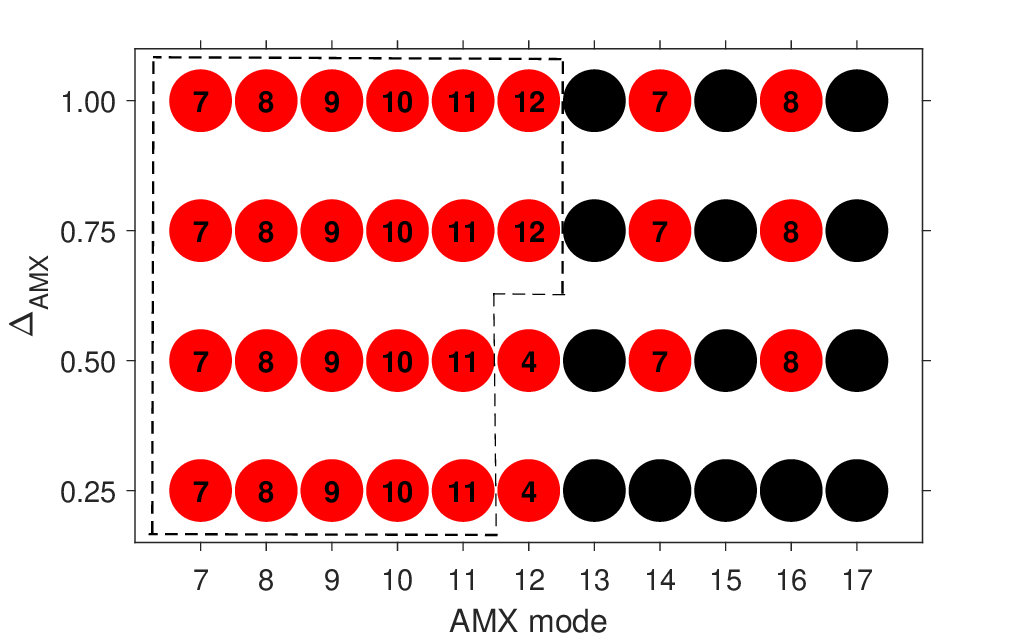}
\vskip -1mm
\caption{Map in the $(\Delta_\mathrm{AMX}, \mathrm{AMX\ mode})$ space showing the occurrence of perfect (red circles) and imperfect (black circles) soliton crystals. For each point, a full detuning scan is performed at $X=3.5$, with the normalized detuning swept from $-0.9$ to $4.9$. Numbers indicate the soliton count for perfect crystals. The dashed contour indicates the region of deterministic PSC formation, where the AMX mode is smaller than or equal to the number of rolls of the first Turing pattern.}
\label{fig6}
\end{figure}
In Sections~\ref{sec_stab}--\ref{sec_pinning}, we have considered some specific cases with different AMX spectral positions and fixed strength. We now systematically investigate the role of both quantities in determining whether perfect or imperfect soliton crystals are formed. To this end, we perform simulations with $\Delta_\mathrm{AMX}$ values from 0.25 to 1 in steps of 0.25, while positioning the AMX at cavity modes between 7 and 17. For each pair $(\Delta_\mathrm{AMX}, \mathrm{AMX\ mode})$, a full detuning sweep is carried out and the resulting soliton crystal state is identified.  For these $\Delta_\mathrm{AMX}$ values the AMX acts as a perturbative effect, not inducing chaotic dynamics during the detuning scan~\cite{Karpov_2019}.

The results are summarized in Fig.~\ref{fig6}, where red circles indicate PSCs and black circles indicate imperfect soliton crystals. We observe that for AMX modes between 7 and 11, PSCs are always formed, with the crystal periodicity matching the AMX mode. Recall that 11 corresponds to the periodicity of the first Turing pattern emerging in the dynamical simulations due to the modulational instability of the CW state. 

The case with the AMX positioned at mode 12 is particularly interesting. A PSC is obtained for the two lowest AMX strengths, but with a 4-FSR spacing, corresponding to a submultiple of the AMX mode, suggesting that the effective AMX pump is not sufficiently strong to promote mode 12 as the dominant comb line. As $\Delta_\mathrm{AMX}$ is increased to $0.75$ or $1$ the AMX becomes sufficiently strong to enforce the formation of a PSC-12 state. Notably, in these two cases the first Turing pattern that forms, i.e. the one arising from the CW destabilization, exhibits a periodicity of 12 rather than 11, which is maintained throughout the entire detuning scan, ultimately acting as a seed for the formation of the PSC-12. This occurs because the AMX enhances the parametric gain at mode 12, thereby favoring the formation of a TP-12 state instead of the usual TP-11 (see Appendix~\ref{MI_sec}, where a rigorous MI analysis in the presence of AMX is presented). The formation of PSC-12 therefore originates from the modification of the modulation-instability gain induced by the AMX.

For AMX modes larger than 12, PSCs form only when the AMX mode is an even number, and the resulting crystal periodicity is half of the AMX mode. This indicates that the spectral line at the AMX mode is present in the spectrum but does not correspond to the dominant comb line due to the weakness of the effective AMX pump term in \eqref{LLE_AMX}. These results show that operating below $X_\mathrm{th,STC}$ is not a sufficient condition for the deterministic formation of PSCs. Moreover, they suggest that when the AMX mode is smaller than or equal to the periodicity of the first Turing pattern that forms — 11 in most of our cases and 12 for a few specific cases — a PSC with the same periodicity as the AMX mode is always obtained. The region where this occurs is highlighted by a dashed contour in Fig.~\ref{fig6}. We note this is consistent with the results reported by Karpov {\sl et al.}~\cite{Karpov_2019}, where the first Turing pattern consists of 23 rolls and the AMX is positioned at mode 15. When the AMX mode exceeds the number of rolls in the first Turing pattern, the effective pump is no longer strong enough to induce a Turing pattern whose periodicity matches the number of sites imposed by the AMX. As a result, PSCs either do not form or emerge with a periodicity lower than the AMX mode. These results hint that if the AMX mode is a prime number larger than the number of rolls of the first Turing pattern, only imperfect soliton crystals can form.

\section{Conclusion} \label{sec_concl}
This theoretical study provides a comprehensive picture of the different mechanisms through which an AMX can induce the formation of soliton crystals. We have shown that the spectral position of the AMX is the key parameter determining the nature of its action. For instance, the AMX can stabilize a PSC that would otherwise be unstable in its absence. This bears similarities to the noise suppression observed with auxiliary-laser injection \cite{moille2025all}.

In most of the considered cases, however, the AMX modifies the periodicity of the Turing patterns, forcing them to match the number of equidistant trapping sites created by the modulated background and ultimately leading to the formation of a PSC.

A clear prediction emerging from our study is that, when the AMX mode position exceeds the number of pulses of the first Turing pattern formed during the detuning scan, PSC formation is no longer deterministic and imperfect soliton crystals can emerge. This can be interpreted as a situation in which the Turing pattern seeding the crystal formation contains fewer pulses than the number of available trapping sites. As a result, the system either forms a PSC with a smaller number of solitons or develops an imperfect crystal. These results therefore show that the AMX can reduce the periodicity of the first Turing pattern formed during the detuning scan, but cannot increase it once the pattern has already formed. This behavior occurs because for large AMX mode numbers, the amplitude of the effective pump associated with the AMX decreases, weakening its influence on the comb dynamics and preventing the comb line at the AMX frequency from becoming the dominant spectral component.

Nevertheless, a mechanism through which the AMX can induce Turing patterns with a periodicity larger than in the AMX-free case is by modifying the parametric gain of the modulation instability. Even in this case, however, the periodicity of the first Turing pattern sets the maximum number of pulses that can be sustained in the resulting soliton crystal.

These results provide a solid theoretical framework for understanding a wide range of experimental observations and for controlling the spectra of frequency combs in microresonators, enabling both perfect and imperfect soliton crystals depending on the target application. 

Our study also shows that stable perfect soliton crystals can form even in the absence of an AMX, although with a reduced periodicity (PSC-7) compared to the Turing pattern seeding the bistable region (TP-9). A detailed investigation of PSC formation without AMX and of its physical origin is left for future work.

We expect that our results will stimulate further experimental investigations on formation and control of soliton crystals, with potential impact on their applications, such as terahertz photonics, data storage, astrophotonics, and optical communications~\cite{Corcoran_2020,Lu_2021,Cole_2017,Karpov_2019}.

\appendix

\section{Modulation instability in presence of an AMX}\label{MI_sec}

\begin{figure}[h!]
\centering
\includegraphics[width=0.5\textwidth]{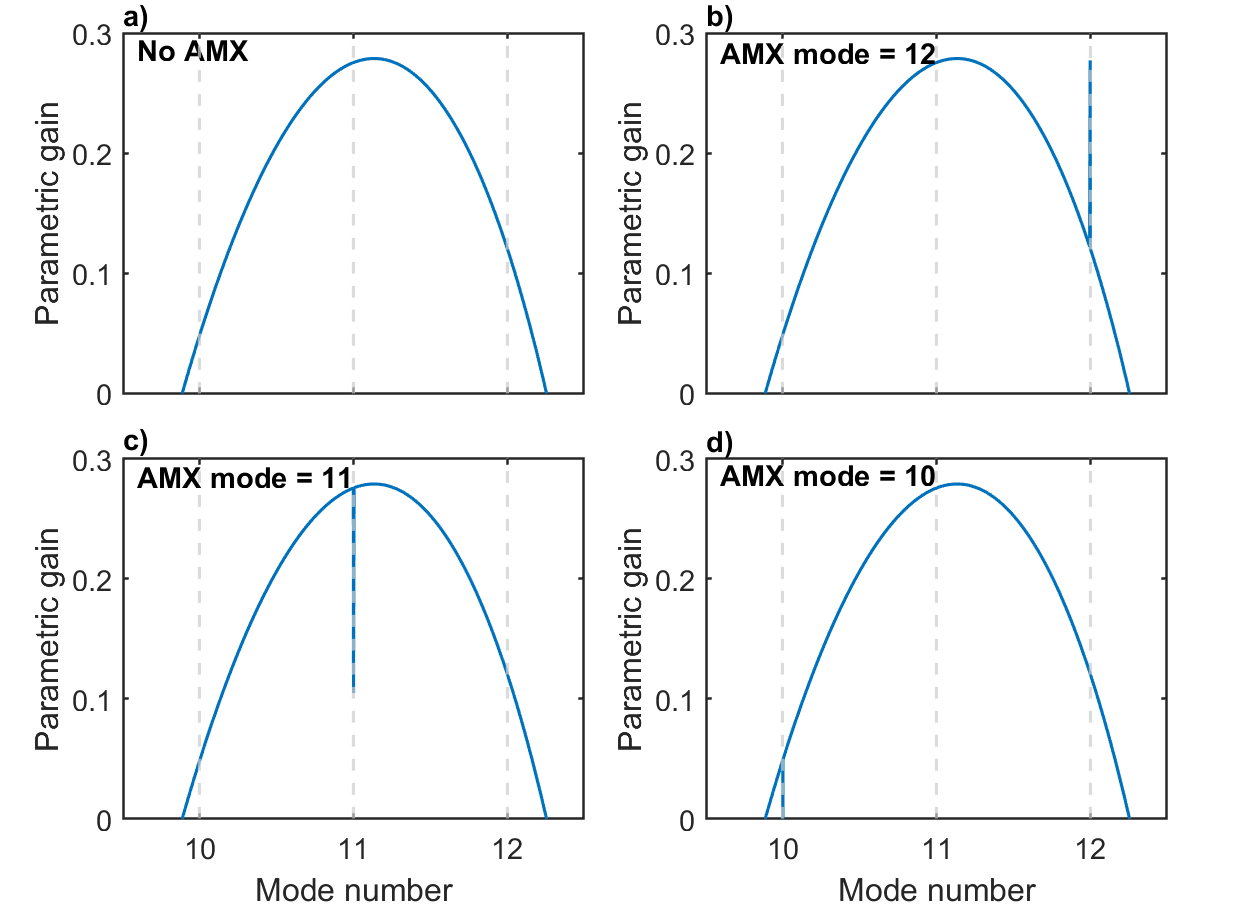}
\vskip -1mm
\caption{Parametric gain at $\Delta=-0.3$ and $X=3.5$ without AMX ($\Delta_\mathrm{AMX}=0$) (a) and with AMX for different AMX positions: mode 12 (b), mode 11 (c), and mode 10 (d), with $\Delta_{\mathrm{AMX}}=0.75$ in all cases. The gray vertical dashed lines indicate the cavity modes.}
\label{fig8}
\end{figure}
To investigate the modulation instability in the presence of an AMX, we consider a CW solution $E_s$ of \eqref{LLE_AMX} and introduce small perturbations oscillating at frequencies $\pm\Omega$:
\begin{equation}
E(\tau,t)=E_s+a(t)e^{i\Omega\tau}+b(t)e^{-i\Omega\tau}.
\label{ansatzMI}
\end{equation}
Substituting ansatz~\eqref{ansatzMI} into the modified LLE~\eqref{LLE_AMX} and linearizing around the CW solution, yields the system
\begin{equation}
\frac{d}{dt}
\begin{pmatrix}
a \\
b^{*}
\end{pmatrix}
=
M
\begin{pmatrix}
a \\
b^{*}
\end{pmatrix},
\end{equation}
with Jacobian matrix
\begin{equation}
\scriptsize
M=
\begin{pmatrix}
-1+K+i\Delta_{\mathrm{AMX}}
\delta(\Omega-\omega_{\mathrm{AMX}})
&
iE_s^2
\\
-i(E_s^*)^2
&
-1-K-i\Delta_{\mathrm{AMX}}
\delta(\Omega+\omega_{\mathrm{AMX}})
\end{pmatrix}.
\label{eq:Jacobian}
\end{equation}
where $K
=
-i\Delta
+
i\eta\Omega^2
+
2i|E_s|^2.$
The parametric gain is then given by
\begin{equation}
G(\Omega)=2\,\mathrm{max}\left[\mathrm{Re}\left(\lambda(\Omega)\right)\right].
\end{equation}
where the $\lambda$ are the eigenvalues of matrix~(\ref{eq:Jacobian}).

If $\Delta_\mathrm{AMX}=0$, the parametric gain of the standard LLE (\eqref{eq:GLLE}) is recovered. Figure~\ref{fig8}(a) shows an example for $\Delta=-0.3$ and $X=3.5$. Figure~\ref{fig8}(b) shows the parametric gain when the AMX is positioned at mode 12, evaluated at $\Delta=-0.3$, corresponding to the onset of modulation instability, and for $\Delta_{\mathrm{AMX}}=0.75$. This parameter set corresponds to one of the cases in which the dynamical simulations exhibit the formation of a TP-12 state (instead of the conventional TP-11), which subsequently evolves into a PSC-12 state (see sixth column of Fig.~\ref{fig6}).
\\
The parametric gain spectrum exhibits a spike-like perturbation at mode 12 induced by the AMX, which increases the gain at this mode to a value comparable to that at mode 11. This clarifies the formation of a TP-12 state observed in the dynamical simulations. Interestingly, when the AMX is instead positioned at mode 11, the AMX-induced perturbation appears as a dip rather than a peak (see Fig.~\ref{fig8}(c)). In this case as well, the parametric gain at mode 12 becomes larger than that at mode 11. Consistently, the dynamical simulations show the destabilization of the CW state in favor of a TP-12 pattern. However, as the detuning is increased, a transition from TP-12 to TP-11 driven by the AMX is observed, after which the TP-11 state remains stable and seeds a PSC with the same periodicity. This explains why the corresponding red circle in Fig.~\ref{fig6} is associated with a PSC-11 state. Finally, when the AMX is positioned at mode 10 while all other parameters are kept unchanged, the parametric gain at mode 10 becomes negative, thereby suppressing the formation of a TP-10 state (Fig.~\ref{fig8}(d)). We verified that in the presence of the AMX, a TP-10 state does not form even when the simulations are initialized with ten equally spaced pulses; instead, the initial condition evolves toward a pattern with 11 rolls.

\section*{Funding}
This research was supported by the Australian Research Council (ARC) Centre of Excellence in Optical Microcombs for Breakthrough Science (project no. CE230100006), funded by the Australian Government. A.F.J.R. is supported by the ARC Discovery Early Career Researcher Award (DE220100509). S.C. is supported by the Royal Society Te Apārangi Marsden fund 23-UOA-053.
\section*{Data availability}
Data underlying the results presented in this Article are not publicly available but may be obtained from the authors upon reasonable request.


\bibliography{AMX_SC}

@article{Gomila2022,
  title = {Role of Eckhaus instability and pattern cracking in ultraslow dynamics of Kerr combs},
  author = {Gomila, Dami\`a and Parra-Rivas, Pedro and Colet, Pere and Coillet, Aur\'elien and Lin, Guoping and Daugey, Thomas and Diallo, Souleymane and Merolla, Jean-Marc and Chembo, Yanne K.},
  journal = {Phys. Rev. A},
  volume = {106},
  issue = {5},
  pages = {053518},
  numpages = {7},
  year = {2022},
  month = {Nov},
  publisher = {American Physical Society},
  doi = {10.1103/PhysRevA.106.053518},
  url = {https://link.aps.org/doi/10.1103/PhysRevA.106.053518}
}

@book{NumRec,
author = {Press, William H. and Teukolsky, Saul A. and Vetterling, William T. and Flannery, Brian P.},
title = {Numerical Recipes 3rd Edition: The Art of Scientific Computing},
year = {2007},
isbn = {0521880688},
publisher = {Cambridge University Press},
address = {USA},
edition = {3},
}

@article{Bartolo:22,
author = {A. Bartolo and N. Vigne and M. Marconi and G. Beaudoin and K. Pantzas and I. Sagnes and G. Huyet and F. Maucher and S. V. Gurevich and J. Javaloyes and A. Garnache and M. Giudici},
journal = {Optica},
number = {12},
pages = {1386--1393},
publisher = {Optica Publishing Group},
title = {Temporal localized Turing patterns in mode-locked semiconductor lasers},
volume = {9},
month = {Dec},
year = {2022},
url = {https://opg.optica.org/optica/abstract.cfm?URI=optica-9-12-1386},
doi = {10.1364/OPTICA.471006},
}

@Article{Perinet2017,
author={P{\'e}rinet, Nicolas
and Verschueren, Nicolas
and Coulibaly, Saliya},
title={Eckhaus instability in the Lugiato-Lefever model},
journal={The European Physical Journal D},
year={2017},
month={Sep},
day={26},
volume={71},
number={9},
pages={243},
issn={1434-6079},
doi={10.1140/epjd/e2017-80078-9},
url={https://doi.org/10.1140/epjd/e2017-80078-9}
}

@article{HerrPRL2014,
  title = {Mode Spectrum and Temporal Soliton Formation in Optical Microresonators},
  author = {Herr, T. and Brasch, V. and Jost, J. D. and Mirgorodskiy, I. and Lihachev, G. and Gorodetsky, M. L. and Kippenberg, T. J.},
  journal = {Phys. Rev. Lett.},
  volume = {113},
  issue = {12},
  pages = {123901},
  numpages = {6},
  year = {2014},
  month = {Sep},
  publisher = {American Physical Society},
  doi = {10.1103/PhysRevLett.113.123901},
  url = {https://link.aps.org/doi/10.1103/PhysRevLett.113.123901}
}

@article{Maggipinto,
  title = {Cavity solitons in semiconductor microresonators:   Existence, stability, and dynamical properties},
  author = {Maggipinto, T. and Brambilla, M. and Harkness, G. K. and Firth, W. J.},
  journal = {Phys. Rev. E},
  volume = {62},
  issue = {6},
  pages = {8726--8739},
  numpages = {0},
  year = {2000},
  month = {Dec},
  publisher = {American Physical Society},
  doi = {10.1103/PhysRevE.62.8726},
  url = {https://link.aps.org/doi/10.1103/PhysRevE.62.8726}
}

@article{ErkintaloPhase,
author = {Miro Erkintalo and Stuart G. Murdoch and Stéphane Coen},
title = {Phase and intensity control of dissipative Kerr cavity solitons},
journal = {Journal of the Royal Society of New Zealand},
volume = {52},
number = {2},
pages = {149--167},
year = {2022},
publisher = {Taylor \& Francis},
doi = {10.1080/03036758.2021.1900296},
URL = {https://doi.org/10.1080/03036758.2021.1900296},
eprint = {https://doi.org/10.1080/03036758.2021.1900296}
}

@article{Wang:17,
author = {Yadong Wang and Fran\c{c}ois Leo and Julien Fatome and Miro Erkintalo and Stuart G. Murdoch and St\'{e}phane Coen},
journal = {Optica},
number = {8},
pages = {855--863},
publisher = {Optica Publishing Group},
title = {Universal mechanism for the binding of temporal cavity solitons},
volume = {4},
month = {Aug},
year = {2017},
url = {https://opg.optica.org/optica/abstract.cfm?URI=optica-4-8-855},
doi = {10.1364/OPTICA.4.000855},
}

@ARTICLE{Corcoran_2020,
   author       = "B.~Corcoran and M.~Tan and X.~Xu and A.~Boes and J.~Wu and T.~G.~Nguyen and S.~T.~Chu and B.~E.~Little and R.~Morandotti and A.~Mitchell and D.~J.~Moss", 
   title        = "Ultra-dense optical data transmission over standard fibre with a single chip source", 
   journal      = "Nat. Comm.", 
   volume       = "11", 
   pages        = "2568", 
   year         = "2020", 
}

@ARTICLE{Lu_2021,
   author       = "Z.~Lu and H.-J.~Chen and W.~Wang and L.~Yao and Y.~Wang and Y.~Yu and B.~E.~Little and S.~T.~Chu and Q.~Gong and W.~Zhao and X.~Yi and Y.-F.~Xiao and W.~Zhang", 
   title        = "Synthesized soliton crystals", 
   journal      = "Nat. Comm.", 
   volume       = "12", 
   pages        = "3179", 
   year         = "2021", 
}

@ARTICLE{Silvestri_2025,
   author       = "C.~Silvestri and Y.~L.~Qiang and K.~Panda and J.~Widjaja and S.~Coen and C.~M.~de Sterke and A.~F.~J.~Runge", 
   title        = "Pure high-order dispersion dissipative Kerr solitons in optical cavities", 
   journal      = "Opt. Lett.", 
   volume       = "50", 
   pages        = "4262-4265", 
   year         = "2025", 
}

@article{LLE_1987,
  title = {Spatial Dissipative Structures in Passive Optical Systems},
  author = {Lugiato, L. A. and Lefever, R.},
  journal = {Phys. Rev. Lett.},
  volume = {58},
  issue = {21},
  pages = {2209--2211},
  numpages = {0},
  year = {1987},
  month = {May},
  publisher = {American Physical Society},
  doi = {10.1103/PhysRevLett.58.2209},
  url = {https://link.aps.org/doi/10.1103/PhysRevLett.58.2209}
}

@ARTICLE{Karpov_2019,
   author       = "M.~Karpov and M.~H.~P.~Pfeiffer and H.~Guo and W.~Weng and J.~Liu and T.~J.~Kippenberg", 
   title        = "Dynamics of soliton crystals in optical microresonators", 
   journal      = "Nat. Phys.", 
   volume       = "15", 
   pages        = "1071–1077", 
   year         = "2019", 
}

@article{Hansch,
  title = {Nobel Lecture: Passion for precision},
  author = {H\"ansch, Theodor W.},
  journal = {Rev. Mod. Phys.},
  volume = {78},
  issue = {4},
  pages = {1297--1309},
  numpages = {0},
  year = {2006},
  month = {Nov},
  publisher = {American Physical Society},
  doi = {10.1103/RevModPhys.78.1297},
  url = {https://link.aps.org/doi/10.1103/RevModPhys.78.1297}
}

@article{BLANES2002313,
title = {Practical symplectic partitioned {R}unge–{K}utta and {R}unge–{K}utta–{N}yström methods},
journal = {Journal of Computational and Applied Mathematics},
volume = {142},
number = {2},
pages = {313-330},
year = {2002},
issn = {0377-0427},
doi = {https://doi.org/10.1016/S0377-0427(01)00492-7},
url = {https://www.sciencedirect.com/science/article/pii/S0377042701004927},
author = {S. Blanes and P.C. Moan}
}

@PREAMBLE{
 "\providecommand{\noopsort}[1]{}" 
 # "\providecommand{\singleletter}[1]{#1}%" 
}

@ARTICLE{Udem_2002,
   author       = "Th.~Udem and R.~Holzwarth and T.~W.~Hänsch", 
   title        = "Optical frequency metrology", 
   journal      = "Nature", 
   volume       = "416", 
   pages        = "233–237", 
   year         = "2002", 
}

@ARTICLE{DelHaye_2007,
   author       = "P.~Del’Haye and A.~Schliesser and O.~Arcizet and T.~Wilken and R.~Holzwarth and T.~J.~Kippenberg", 
   title        = "Optical frequency comb generation from a monolithic microresonator", 
   journal      = "Nature", 
   volume       = "450", 
   pages        = "1214–1217", 
   year         = "2007", 
}

@ARTICLE{Herr_2014,
   author       = "T.~Herr and V.~Brasch and J.~D.~Jost and C.~Y.~Wang and N.~M.~Kondratiev and M.~L.~Gorodetsky and T.~J.~Kippenberg",
   title        = "Temporal solitons in optical microresonators", 
   journal      = "Nat. Photonics", 
   volume       = "8", 
   pages        = "145–152", 
   year         = "2014", 
}

@ARTICLE{Kippenberg_2018,
   author       = "T.~J.~Kippenberg and A.~L.~Gaeta and M.~Lipson and M.~L.~Gorodetsky",
   title        = "Dissipative {K}err solitons in optical microresonators",
   journal      = "Science",
   volume       = "361", 
   pages        = "eaan8083",
   year         = "2018",
}

@ARTICLE{Coen_2013,
   author       = "S.~Coen and H.~G.~Randle and T.~Sylvestre and M.~Erkintalo", 
   title        = "Modeling of octave-spanning {K}err frequency combs using a generalized mean-field Lugiato–Lefever model", 
   journal      = "Opt. Lett.", 
   volume       = "38", 
   pages        = "37-39", 
   year         = "2013", 
}

@ARTICLE{Pasquazi_2018,
   author       = "A.~Pasquazi and M.~Peccianti and L.~Razzari and D.~J.~Moss and S.~Coen and M.~Erkintalo and Y.~K.~Chembo and T.~Hansson and S.~Wabnitz and P.~Del’Haye and X.~Xue and A.~M.~Weiner and R.~Morandotti", 
   title        = "Micro-combs: A novel generation of optical sources", 
   journal      = "Phys. Rep.", 
   volume       = "729", 
   pages        = "1-81", 
   year         = "2018", 
}

@ARTICLE{Leo_2013,
   author       = "F.~Leo and L.~Gelens and P.~Emplit and M.~Haelterman and S.~Coen",
   title        = "Dynamics of one-dimensional {K}err cavity solitons", 
   journal      = "Opt. Express", 
   volume       = "21", 
   pages        = "9180-9191", 
   year         = "2013", 
}

@ARTICLE{Cole_2017,
   author       = "D.~C.~Cole and E.~S.~Lamb and P.~Del’Haye and S.~A.~Diddams and S.~B.~Papp",
   title        = "Soliton crystals in Kerr resonators", 
   journal      = "Nat. Photonics", 
   volume       = "11", 
   pages        = "671–676", 
   year         = "2017", 
}

@ARTICLE{Jang_2015,
   author       = "J.~K.~Jang and M.~Erkintalo and S.~G.~Murdoch and S.~Coen", 
   title        = "Writing and erasing of temporal cavity solitons by direct phase modulation of the cavity driving field", 
   journal      = "Opt. Lett.", 
   volume       = "40", 
   pages        = "4755-4758", 
   year         = "2015", 
}

@ARTICLE{Barland_2002,
   author       = "S.~Barland and J.~R.~Tredicce and M.~Brambilla and L.~A.~Lugiato and S.~Balle and M.~Giudici and T.~Maggipinto and L.~Spinelli and G.~Tissoni and T.~Knödl and M.~Miller and R.~Jäger", 
   title        = "Cavity solitons as pixels in semiconductor microcavities", 
   journal      = "Nature", 
   volume       = "419", 
   pages        = "699-702", 
   year         = "2002", 
}

@ARTICLE{Coen2_2013,
   author       = "S.~Coen and M.~Erkintalo", 
   title        = "Universal scaling laws of Kerr frequency combs", 
   journal      = "Opt. Lett.", 
   volume       = "38", 
   pages        = "1790-1792", 
   year         = "2013", 
}

@article{Sun2023Mar,
	author = {Sun, Yang and Wu, Jiayang and Wu, Jiayang and Tan, Mengxi and Xu, Xingyuan and Li, Yang and Morandotti, Roberto and Mitchell, Arnan and Moss, David J. and Moss, David J.},
	title = {{Applications of optical microcombs}},
	journal = {Adv. Opt. Photonics},
	volume = {15},
	number = {1},
	pages = {86--175},
	year = {2023},
	month = mar,
	issn = {1943-8206},
	publisher = {Optica Publishing Group},
	doi = {10.1364/AOP.470264}
}

@article{Qi2019Sep,
	author = {Qi, Zhen and Wang, Shaokang and Jaramillo-Villegas, Jos{\ifmmode\acute{e}\else\'{e}\fi} and Qi, Minghao and Weiner, Andrew M. and D{'}Aguanno, Giuseppe and Carruthers, Thomas F. and Menyuk, Curtis R.},
	title = {{Dissipative cnoidal waves (Turing rolls) and the soliton limit in microring resonators}},
	journal = {Optica},
	volume = {6},
	number = {9},
	pages = {1220--1232},
	year = {2019},
	month = sep,
	issn = {2334-2536},
	publisher = {Optica Publishing Group},
	doi = {10.1364/OPTICA.6.001220}
}

@article{Wang2018May,
	author = {Wang, Weiqiang and Lu, Zhizhou and Zhang, Wenfu and Chu, Sai T. and Little, Brent E. and Wang, Leiran and Xie, Xiaoping and Liu, Mulong and Yang, Qinghua and Wang, Lei and Zhao, Jianguo and Wang, Guoxi and Sun, Qibing and Liu, Yuanshan and Wang, Yishan and Zhao, Wei},
	title = {{Robust soliton crystals in a thermally controlled microresonator}},
	journal = {Opt. Lett.},
	volume = {43},
	number = {9},
	pages = {2002--2005},
	year = {2018},
	month = may,
	issn = {1539-4794},
	publisher = {Optica Publishing Group},
	doi = {10.1364/OL.43.002002}
}

@article{taheri2017optical,
  title={Optical lattice trap for Kerr solitons},
  author={Taheri, Hossein and Matsko, Andrey B and Maleki, Lute},
  journal={The European Physical Journal D},
  volume={71},
  number={6},
  pages={153},
  year={2017},
  publisher={Springer}
}

@article{moille2025all,
  title={All-optical noise quenching of an integrated frequency comb},
  author={Moille, Gr{\'e}gory and Shandilya, Pradyoth and Stone, Jordan and Menyuk, Curtis and Srinivasan, Kartik},
  journal={Optica},
  volume={12},
  number={7},
  pages={1020--1030},
  year={2025},
  publisher={Optica Publishing Group}
}

\end{document}